\documentstyle[11pt,newpasp,twoside,epsf]{article}
\def\edcomment#1{\iffalse\marginpar{\raggedright\sl#1\/}\else\relax\fi}
\reversemarginpar

\begin{document}
\title{A search for nearby young stars among the flare stars}
 \author{Brigitte K\"onig}
\author{Ralph Neuh\"auser}
\affil{MPI f\"ur extraterrestrische Physik, Giessenbachstra\ss e 1,
D-85740 Garching, Germany}
\author{Valeri Hambaryan}
\affil{Astrophysikalisches Institut Potsdam, An der Sternwarte 16, D-14482
Potsdam, Germany} 

\begin{abstract}
Flare stars were discovered in the late 1940s in the solar
vicinity and were named UV Cet-type variables (classical FSs).
Among the FSs within 100\,pc we search for young stars. For the search we
take spectra with sufficient resolution to resolve Lithium at 6707\,\AA~and
Calcium at 6718\,\AA~of all the stars. The real young stars are prime targets
for the search of extra-solar planets by direct imaging.
\end{abstract}

\section{Introduction}
According to the recent definition given by Gershberg et al. (1999), these
stars are on the lower part of the MS and show activity similar to the sun
(sporadic flares, dark spots, variable emission from the chromosphere and the
corona, radio, X-ray and UV outbursts).

The UV Cet-type stars are relatively young (possibly zero-age or pre-MS stars)
with ages $\sim 100$\,Myrs or younger. They may be former members of recently
dispersed T or OB associations or may be ejected stars from associations by a
three body encounter (like run-away T Tauri stars). A local dispersed association does
not appear unlikely: The translucent
high-latitude cloud MBM 12 with ongoing star formation is located at $\sim
65$\,pc (Hearty et al. 2000) and the TW Hya association at $\sim 55$\,pc, which
has dispersed its gas and dust (Webb et al. 1999) are already known. Hence,
there are indeed young T associations within $\sim 100$\, pc. A few isolated
young nearby stars are also known, e.g. GJ~182 at 27\,pc and
20\,Myrs.

\section{The observations and first results}
We observe all these stars with FOCES and FEROS, the high
resolution echelle spectrograph of the Calar Alto Observatory in Spain
and at ESO La Silla Observatory in Chile, respectively.
A first reduction of the spectra of some stars shows the existence of strong
Lithium absorption line at 6708\,\AA. As an example, 
HIP~71631 shows strong Lithium absorption at 6707\,\AA with $\rm
W_{\lambda}({\rm Li}) = 0.213$\,\AA, stronger than
Calcium at 6718\,\AA. This star has been classified as G0.Ve. The parallax
given by HIPPARCOS measurement is $29.46\pm0.61$\,mas which corresponds to a
distance of $33.9\pm0.7$\,pc. The V-magnitude is
$7.613\pm0.011$\,mag. According to the HR-diagramm and the D'Antona \&
Mazzitelli (1994) tracks and isochrones the star has $1.1 \pm 0.1 \rm
M_{\odot}$~and an age of $5 \pm 2.4 \cdot 10^7$~years. 

\begin{figure}
\plotfiddle{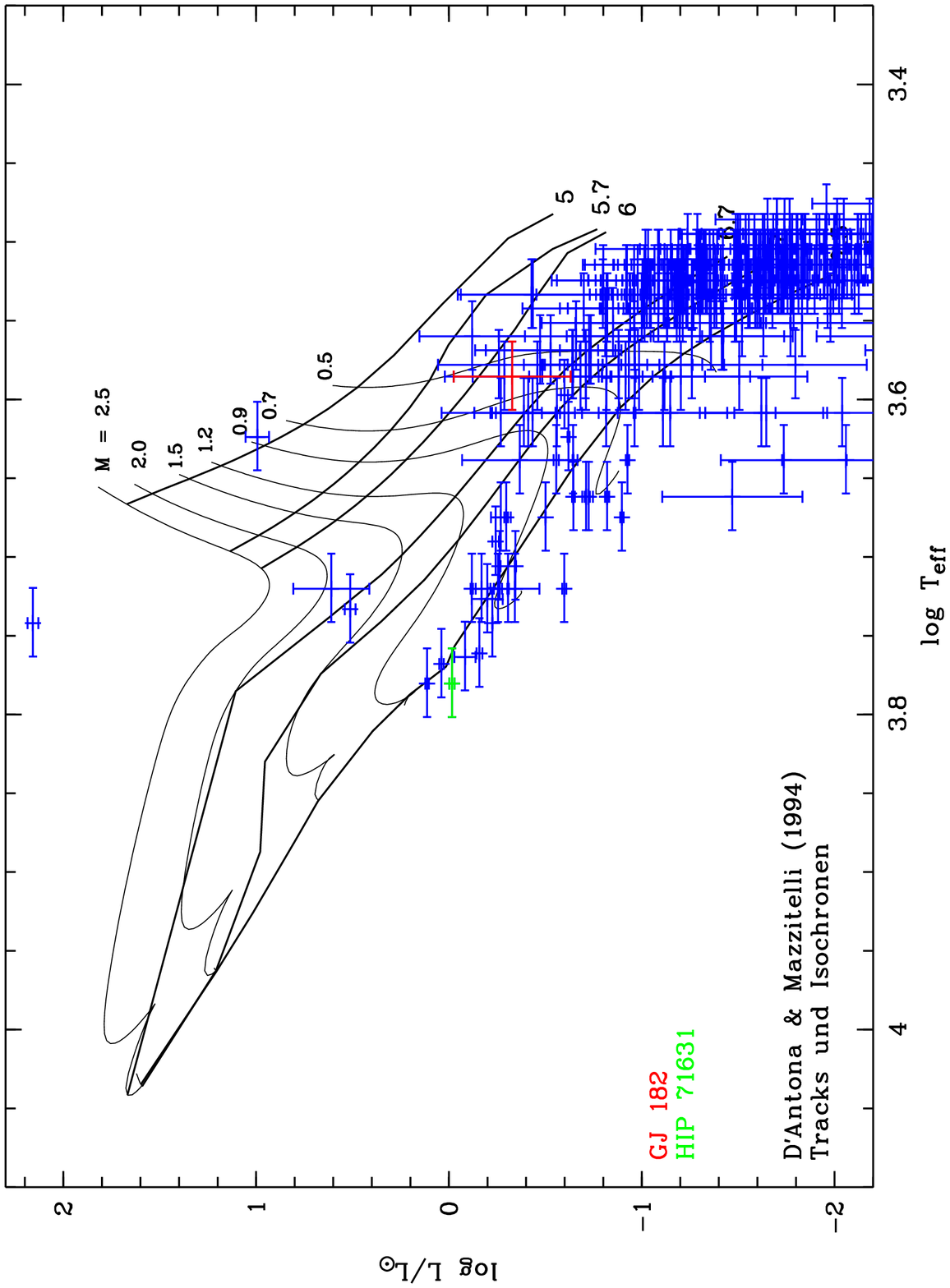}{\textwidth}{270}{30}{30}{-225}{400}

\plotfiddle{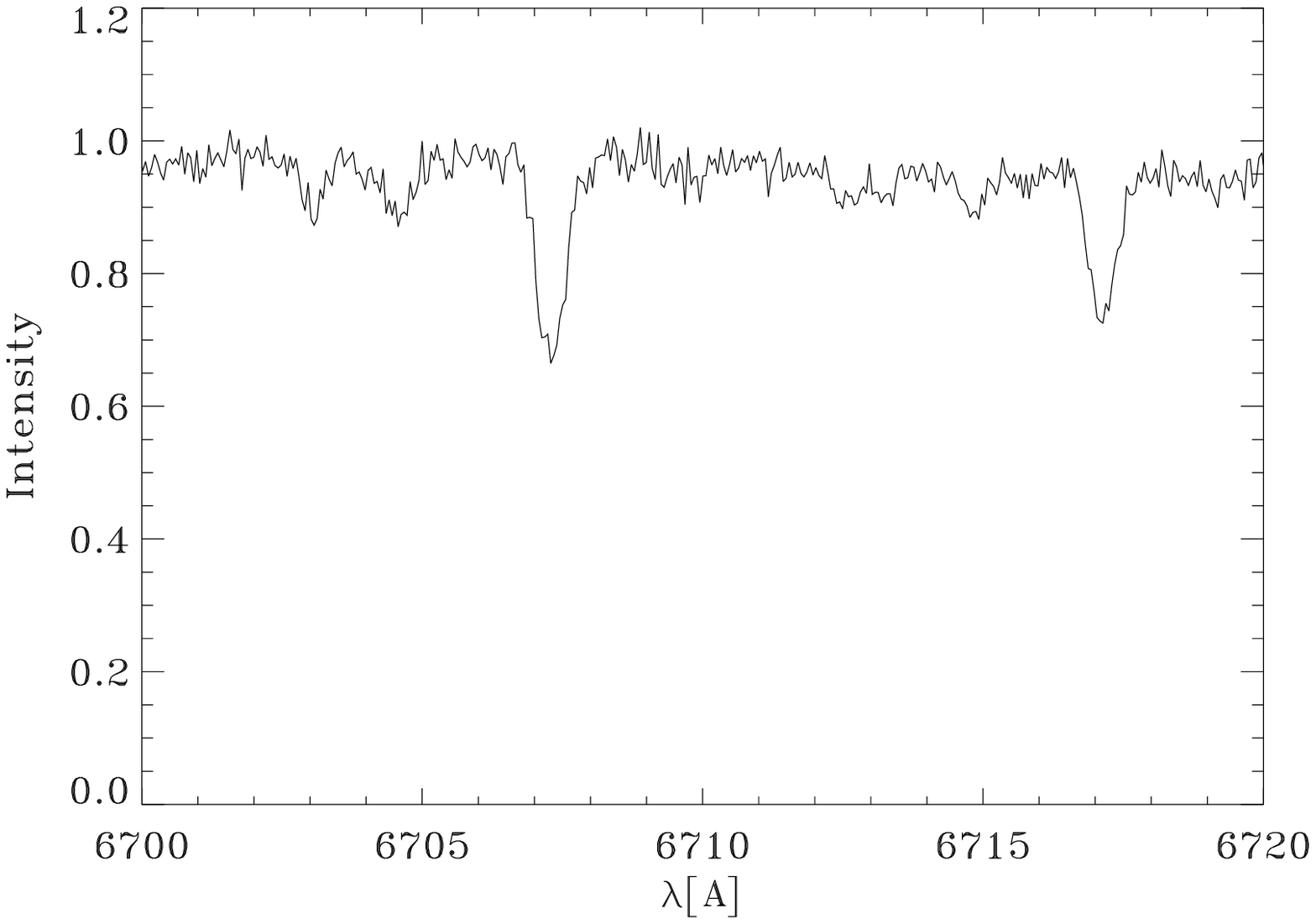}{\textwidth}{0}{40}{48}{-20}{440}
\vspace{-22cm}
\caption{Left: D'Antona \& Mazzitelli (1994) tracks and isochrones together
with all the stars of the sample. Right: An example for a high resolution
spectrum of the star HIP~71531 where we found Lithium at 6708\,\AA~stronger
than Calcium at 6718\,\AA.}
\end{figure}

\section{Future work}
The high resolution echelle spectra enables us to study the stellar properties
of the stars in detail. Especially we can measure radial
velocity, temperature, metallicity, gravity and rotational
velocity ($v \sin{i}$) to distinguish between pre-MS-dwarfs and
Post-MS-giants. Stars in the sample which appear young will be proposed for
VLBI observations to spatially resolve possible coronal loops and they are
prime targets for direct imaging of extra-solar planets in the IR.

\end{document}